\documentstyle[aps,preprint,tighten,floats]{revtex}

\input epsf

\def \MSbar {\vbox{\hrule\kern 1pt\hbox{\rm MS}}}
\def \GeV { {\ \rm GeV} }

\begin{document}

\draft

\title{Factorization scheme dependence \\ 
of the NLO inclusive jet cross section}

\author{Parvez Anandam and Davison E. Soper}

\address{Institute of Theoretical Science\\
University of Oregon, Eugene, OR 97403}

\date{13 December 1999}

\maketitle

\begin{abstract}
We study the factorization scheme dependence of the next-to-leading
order inclusive one jet cross section $d\sigma/dE_T$. The scheme is
varied parametrically along the direction that transforms the
$\overline{\mathrm{MS}}$ scheme to the DIS scheme: we introduce a
parameter $\lambda$ such that $\lambda=0$ is
$\overline{\mathrm{MS}}$ and $\lambda=1$ is DIS. The factorization
scale $\mu$ is also varied. We observe a change of $\pm9\%$ in the
cross section for $E_T =50\ \mathrm{GeV}$ when $\mu$ and $\lambda$ are
varied in the range $-2 \le \lambda \le 2$, $E_T/8 \le \mu \le 2
E_T$. This grows to $\pm 32\%$ for $E_T = 400\ \mathrm{GeV}$.
\end{abstract}


Calculations in perturbative QCD involve conventions. For example, one
conventionally uses the \MSbar\ scheme for renormalization. Within the
\MSbar\ scheme, one needs to specify a renormalization scale $\mu_{\rm
UV}$. If one does a calculation at next-to-next-to-leading order
(NNLO), it is possible to vary the renormalization scheme by setting
the third coefficient in the $\beta$ function, $b_3$. It is useful to
study the dependence of the calculated cross sections on the chosen
conventions. Thus, with a NNLO calculation of the total cross section
for $e^+ + e^- \to hadrons$, one can examine the dependence of the
cross section on $\mu_{\rm UV}$ and $b_3$ \cite{DSLS}. Studies like
this have a dual purpose. First, we learn about the extent to which
our arbitrary choices affect the results. Second, we obtain an
estimate, or at least an approximate lower bound, on the theoretical
uncertainty in the calculated order $\alpha_s^N$ cross section arising
from uncalculated terms of order $\alpha_s^{N+1}$ and higher. This is
because among the uncalculated terms are contributions containing
factors of $\ln(\mu_{\rm UV})$ and $b_3$ that cancel the $\mu_{\rm
UV}$ and $b_3$ dependence in the calculated terms. We cannot find all
of the $\alpha_s^{N+1}$ terms (since we lack the calculation), but at
least we know the size of some of them.

In cross sections involving initial state hadrons, the theoretical
formula contains parton distribution functions as factors. The
calculated result at order $\alpha_s^N$ depends on the convention we
use for collinear factorization. Put another way, the result depends
on the definition of the parton distribution functions. In particular,
the result depends on the factorization scale $\mu_{\rm coll}$ (the
scale that appears in the parton distributions $f_{a/A}(x,\mu_{\rm
coll})$).  In NLO calculations of hard cross sections with initial
state hadrons, it is common to examine the dependence of the result on
$\mu_{\rm coll}$.

In this paper, we examine a bigger space of convention dependence. The
dependence of the parton distribution functions on the factorization
scale has the form
\begin{eqnarray}
f_{a/A}(x,\tilde{\mu}) &=& f_{a/A}(x,\mu)
\nonumber\\
& & \mbox{} + {\alpha_s(\mu)\over 2\pi}\,
\ln(\tilde{\mu}^2 / \mu^2 )
\int_x^1 {d\xi\over \xi} \sum_b K^{(0)}_{ab}( x / \xi )\,
f_{b/A}\!\left(\xi,\mu \right)
\nonumber\\
& & \mbox{} + {\cal O}(\alpha_s^2) .
\end{eqnarray}
Here $K^{(0)}_{ab}(\xi)$ is the lowest order Altarelli-Parisi kernel.
If we like, we can redefine the parton distributions according to
\begin{eqnarray}
\tilde{f}_{a/A}(x,\mu) &=& f_{a/A}(x,\mu)
\nonumber\\
& & \mbox{} + {\alpha_s(\mu)\over 2\pi}\, \sum_{J = 0}^N
\lambda_J
\int_x^1 {d\xi\over \xi} \sum_b K^{(J)}_{ab} ( x / \xi )\,
f_{b/A}\!\left(\xi,\mu\right).
\label{pdfdef}
\end{eqnarray}
Here $K^{(0)}_{ab}(\xi)$ is still the Altarelli-Parisi kernel, so that
if we put $\lambda_0 = \ln\left(\tilde{\mu}^2 / \mu^2\right)$
then the $J=0$ term amounts to a change of scale, considered at lowest
order in $\alpha_s$. The essential new ingredient is that we now have
added more terms with kernels $K^{(J)}_{ab}(\xi)$, which can be
anything we please.

Let us suppose that we are calculating a cross section $\sigma$. If we
calculate the hard scattering (partonic) cross section at leading
order, then the order $\alpha_s$ change in definition of the parton
distributions caused by varying one of the $\lambda_J$ will lead to a
relative order $\alpha_s$ change in the cross section,
\begin{equation}
{ d \sigma \over d\lambda_L} \propto \sigma \times \alpha_s
\hskip 1 cm ({\rm LO}).
\end{equation}
If we calculate the hard scattering cross section at ${\rm N}^k{\rm
LO}$, then the ${\rm N}^k{\rm LO}$ contribution to the hard scattering
functions will contain contributions proportional to the $\lambda_J$
raised to powers up to $k$, so that the $\lambda_J$ dependence is
partially canceled and
\begin{equation}
{ d \sigma \over d\lambda_L} \propto \sigma \times
\alpha_s^{k+1}
\hskip 1 cm ({\rm N}^k{\rm LO}).
\end{equation}
Suppose that we have a calculation at ${\rm NLO} = {\rm N}^1{\rm LO}$.
Then the change in the cross section induced by changing some of the
$\lambda_J$ provides an indication of uncalculated contributions to
$\sigma$ that are suppressed by a factor $\alpha_s^2$. Looked at
another way, changing some of the $\lambda_J$ provides an indication
of the sensitivity of the calculation to the arbitrary choice of
factorization convention.

Note that the question of the dependence on the factorization convention
is analogous to the question of the dependence on the renormalization
convention. However, in the case of renormalization we are really
varying one quantity, $\alpha_s$, so that there is a one parameter
space of conventions to explore if we work at first order, $\Delta
\alpha_s \propto \alpha_s \times \alpha_s$. In the case of
factorization, we are varying functions, the parton distribution
functions, so that there are an infinite number of parameters to
examine at first order, $\Delta f \propto f \times \alpha_s$.

In this paper, we restrict the parameter space by taking $N = 1$, with
$K^{(1)}$ being the kernel such that $\lambda_1 = 1$ corresponds to
changing from the \MSbar\ scheme to the DIS scheme. We examine the
cross section $d\sigma/dE_T$ for $p \bar p \to {\rm jet} + X$
calculated at NLO and look at how this cross section depends on
$\lambda_0$ and $\lambda_1$.

In our calculation, we eliminate the $J=0$ term in Eq.~(\ref{pdfdef})
and instead change the scale $\mu$ in the parton distributions
directly, using the parton evolution that is incorporated in the
CTEQ4M set of parton distributions.\footnote{The CTEQ4M parton
distributions incorporate evolution with both $\alpha_s$ and
$\alpha_s^2$ terms in the evolution kernel.}  This differs from using
the $J=0$ term at fixed $\mu$ by terms of order $\alpha_s^2$. Thus, we
have a two dimensional parameter space of conventions to explore,
specified by $\mu \equiv \mu_{\mathrm{coll}}$ and $\lambda \equiv
\lambda_1$.

Using the program \cite{EKS}, we computed the one jet inclusive cross
section $d\sigma/d E_T$, where $E_T$ is the transverse energy of the
jet. Suitable modifications for implementing changes in scheme were
made. Jet definitions followed the Snowmass cone algorithm with the
cone radius $R = 0.7$. We averaged over the rapidity $y$ in the range
$0.1 \le |y| \le 0.7$. The renormalization scale was fixed at
$\mu_{\mathrm{UV}} = E_T/2$ to limit the number of variables. The
factorization scale $\mu$ was varied in the range $E_T/8 \le \mu \le 2
E_T$. The factorization scheme parameter $\lambda$ was varied in the
range $-2 \le \lambda \le 2$, with $\lambda=0$ corresponding to the
$\overline{\mathrm{MS}}$ scheme and $\lambda=1$ to the DIS scheme.

We calculate the DIS parton distribution functions from the
$\overline{\mathrm{MS}}$ ones, with CTEQ4M as our standard starting
distribution:
\begin{eqnarray}
\tilde{f}_a(x,\mu,\lambda) = f_a^{\overline{MS}}(x,\mu) + \lambda
\frac{\alpha_s(\mu)}{2\pi}\int_x^1 \frac{d\xi}{\xi} \sum_b
K^{(1)}_{ab}( x/ \xi) f_b^{\overline{MS}}(\xi,\mu)
.
\label{transform}
\end{eqnarray}
The kernel $K^{(1)}$ consists of the usual DIS functions
\cite{handbook},
\begin{eqnarray}
K^{(1)}_{qq}(x) = - K^{(1)}_{gq}(x) & = & C_F \left[ 2\left(
\frac{\ln(1-x)}{1-x} \right)_+ - \frac{3}{2} \left( \frac{1}{1-x} \right)_+
- (1+x)\ln(1-x) \right. \nonumber\\ 
& & - \left. \frac{1+x^2}{1-x} \ln x + 3 + 2x - \left(
\frac{\pi^2}{3} + \frac{9}{2} \right) \delta(1-x) \right] , \nonumber \\
K^{(1)}_{qg}(x) = - \frac{1}{2 n_f} K^{(1)}_{gg}(x) & = & T_F \left[
\left( (1-x)^2 + x^2 \right) \ln \! \left( \frac{1-x}{x} \right) -8x^2
+8x - 1 \right],
\end{eqnarray}
with $q = u,\bar{u},d,\bar{d},$ \ldots \ Here, $C_F = 4/3$, $T_F = 1/2$ 
and the other $K^{(1)}_{ab}$ are equal to zero.

The jet cross section is originally defined as
\begin{eqnarray}
\sigma = 
\left(\frac{\alpha_s}{2\pi}\right)^2
\int_0^1 dx_A \int_0^1 dx_B \sum_{a,b} f_a(x_A) f_b(x_B)
\left[ \hat{\sigma}_0(a,b,x_A, x_B) 
+ \frac{\alpha_s}{2\pi} \,
\hat{\sigma}_1(a,b,x_A,x_B) \right] ,
\end{eqnarray}
where $\hat{\sigma}_0$ is the leading-order (or Born) term, and
$\hat{\sigma}_1$ is the next-to-leading-order term. We can selectively
turn on or off the NLO term in the cross section. We reexpress this
cross section in terms of the new parton distributions $\tilde{f}(x)$,
dropping contributions suppressed by two powers of $\alpha_s$ compared 
to the Born cross section,
\begin{eqnarray}
\sigma & = & 
\left(\frac{\alpha_s}{2\pi}\right)^2
\int_0^1 dx_A \int_0^1 dx_B \sum_{a,b} \biggl\{  
\tilde{f}_a(x_A) \tilde{f}_b(x_B) \left[ 
\hat{\sigma}_0(a,b,x_A, x_B) + 
\frac{\alpha_s}{2\pi}\,
\hat{\sigma}_1(a,b,x_A, x_B) \right]  
\nonumber\\
& & \mbox{} - \frac{\alpha_s}{2\pi} \sum_{J=0}^{1} \lambda_J \int_x^1
\frac{d\xi}{\xi} \sum_{c}  \left[K^{(J)}_{ac}(x_A / \xi) 
\tilde{f}_c(\xi)\tilde{f}_b(x_B) 
+ \tilde{f}_a(x_A) K^{(J)}_{bc}( x_B / \xi ) 
\tilde{f}_c(\xi)\right] 
\nonumber\\
&&
\hskip 4 cm\times 
\hat{\sigma}_0(a,b,x_A, x_B) \biggr\} ,
\end{eqnarray}
where $\lambda_0 = \ln(4\mu^2/E_T^2)$ and $\alpha_s =
\alpha_s(\mu_{\mathrm{UV}})$. The added terms modify $\hat{\sigma}_1$
from the \textit{standard} $\overline{\textit{MS\,}}$ \textit{scheme},
that is the \MSbar\ scheme where $\mu=E_T/2$. (Of course the $J=0$ term
is already included in the standard program \cite{EKS}.) Changing both
parton distribution and jet cross section definitions in this way
exactly cancels out contributions of order $\alpha_s$ to the change in
$\sigma$. What remains are the order $\alpha_s^2$ contributions and the
extent of the change in the cross section gives us an estimate of
these NNLO terms.

\begin{figure}[h]
\begin{tabular}{ll}	
\epsfxsize=2.8in \epsfbox{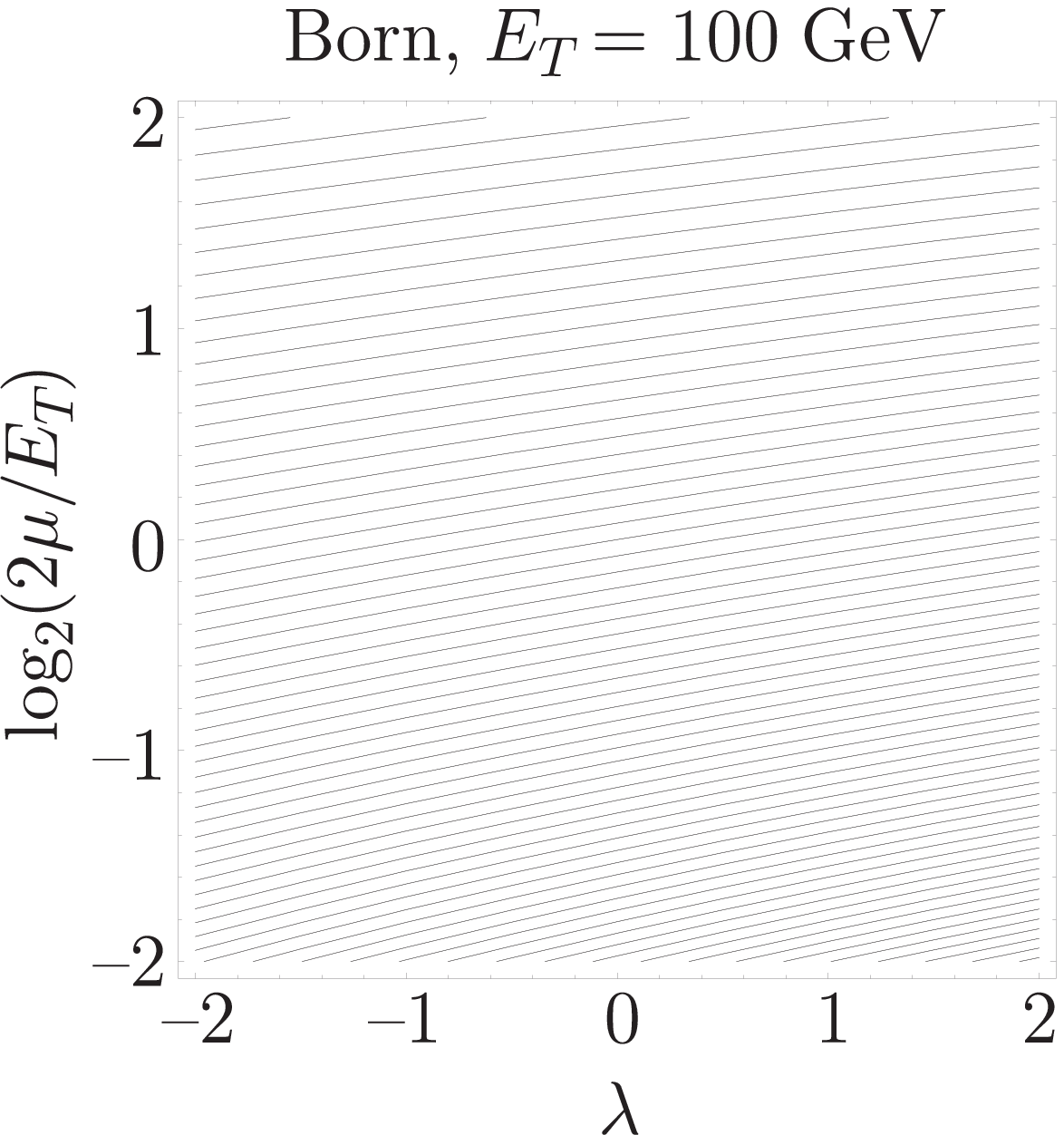} 
& \hspace{0.4in} \epsfxsize = 2.8in \epsfbox{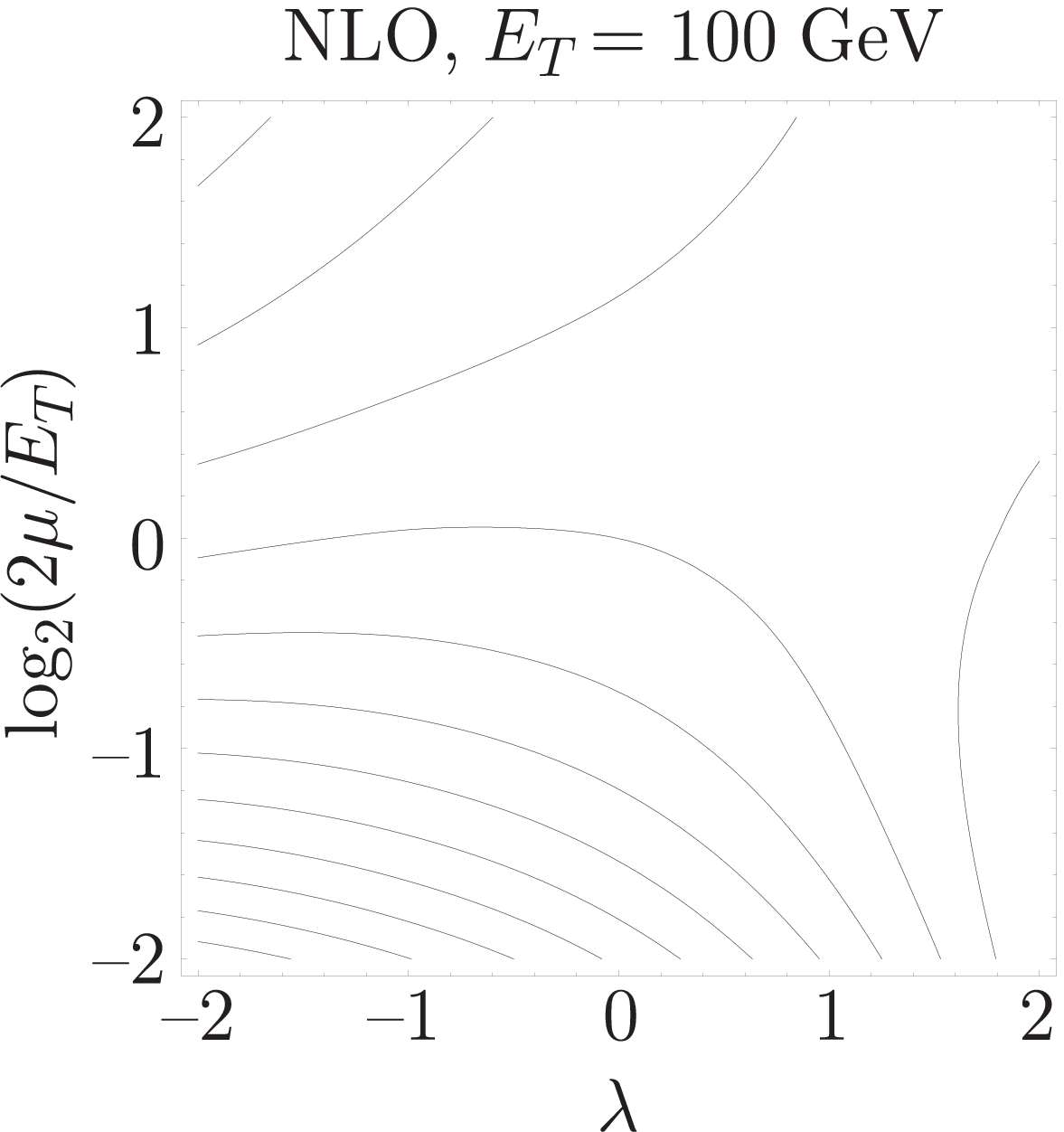}\\
\hspace{1.5in} (a) & \hspace{1.95in} (b)
\end{tabular}
\vspace{0.1in}
\caption{Contour plots of the Born and NLO jet cross section 
at $E_T = \mathrm{100 \ GeV} $ with 1\% contour lines}
\label{pt100contourplots}
\end{figure}

We first calculate the jet cross section $d\sigma/dE_T$ at Born level
at the transverse energy $E_T = \mathrm{100
\GeV}$. Fig.~\ref{pt100contourplots}(a) shows a contour plot of the
fractional difference of this computed cross section from the standard
$\overline{\mathrm{MS}}$ cross section when $\lambda$ and $\mu$ are
varied. The contours represent 1\% changes in the cross section. The
Born level cross section is very sensitive to changes in conventions;
it varies by 40\% from its standard value in the region shown in the
plot.

In Fig.~\ref{pt100contourplots}(b), we show the $\mu$ and $\lambda$
dependence of the cross section calculated at NLO\@. The NLO terms
cancel out most of the convention dependence and a saddle region is
observed. The maximal change from the standard cross section is only
8\%. There is a rather broad region in the parameter space under
consideration where the NLO cross section changes little.

\begin{figure}[h]
\begin{tabular}{ll}	
\epsfxsize=2.8in \epsfbox{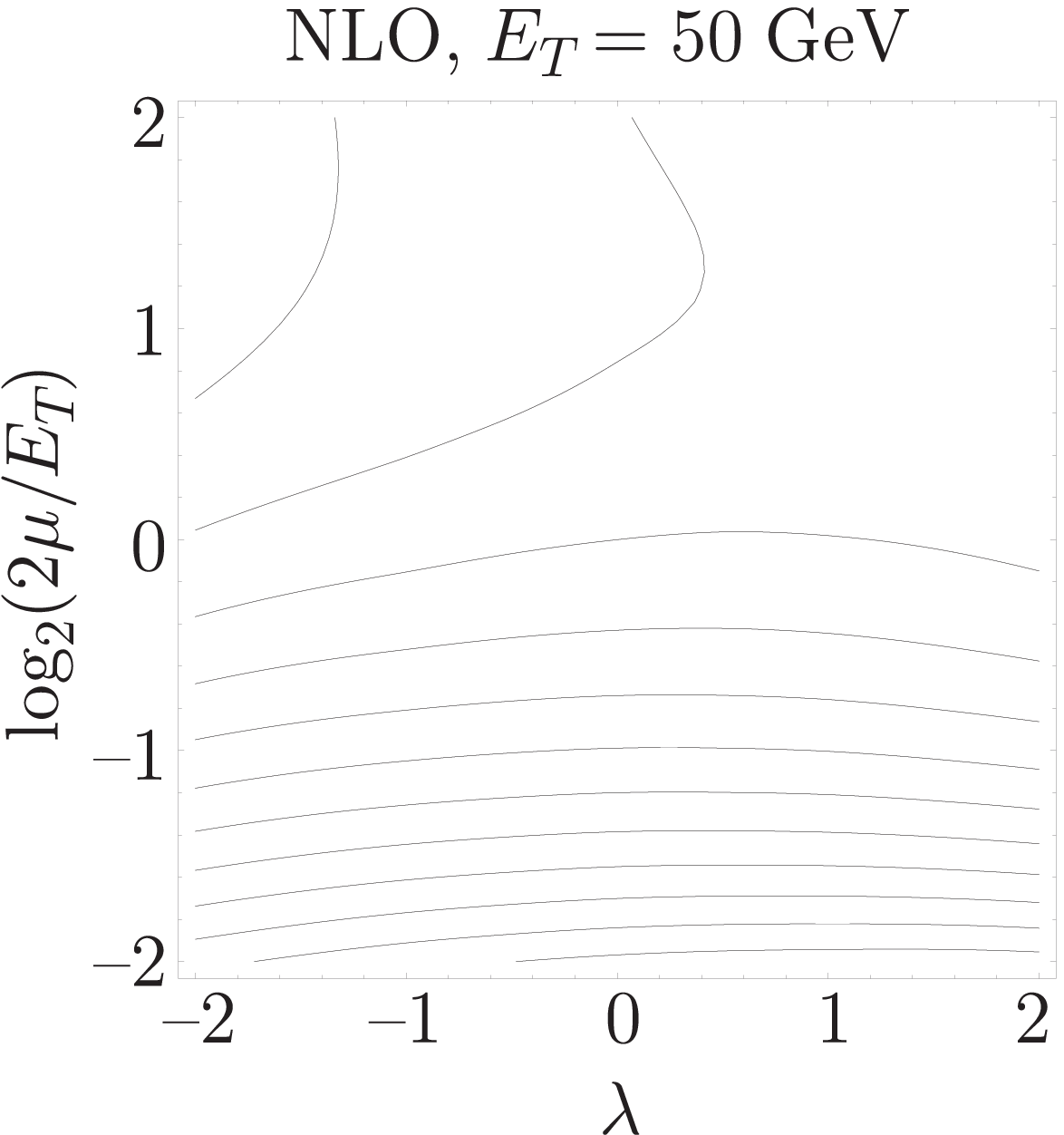}
& \hspace{0.4in} \epsfxsize = 2.8in \epsfbox{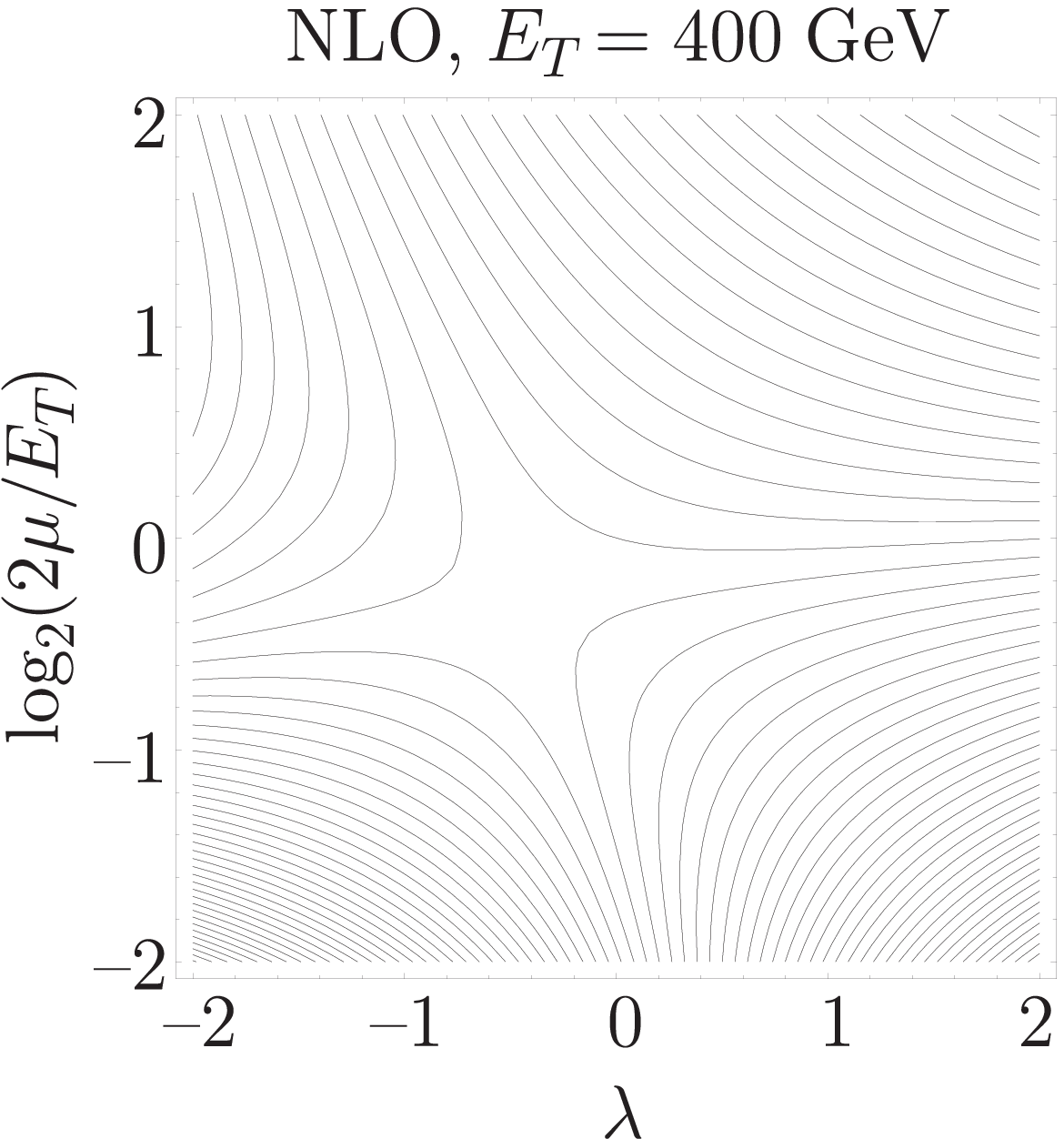}\\
\hspace{1.5in} (a) & \hspace{1.95in} (b)
\end{tabular}
\vspace{0.1in}
\caption{Contour plots of the NLO jet cross section at 
$E_T = \mathrm{50 \ GeV} $ and $E_T = \mathrm{400 \ GeV}$ with 1\%
contour lines}
\label{nlocontourplots}
\end{figure}

In Fig.~\ref{nlocontourplots}, we investigate how this result depends
on $E_T$ in the range $\mathrm{50 \ GeV} \le E_T \le \mathrm{400 \
GeV}$. The sensitivities of both the Born cross section (not shown)
and the NLO cross section to $\mu$ and $\lambda$ increase with
increasing $E_T$. However, the Born cross section is always far more
sensitive to the $\lambda$ and $\mu$ parameters than the NLO cross
sections. At NLO, saddle regions are observed for all values of $E_T$
shown, although the saddle point is not always at the same position in
the $\lambda \, \mu$ plane.

We can offer two comments about the results shown in
Figs.~\ref{pt100contourplots} and \ref{nlocontourplots}. First, the
choice $\mu = E_T/2$ is quite felicitous: with this choice, the
dependence on $\lambda$ is almost nil. Second, there is a lot more
convention dependence at $E_T = 400\GeV$ than at $E_T = 100\GeV$ or
$E_T = 50\GeV$. (One misses seeing this if one looks only at the $\mu$
dependence at $\lambda = 0$. Instead, one needs to vary $\lambda$ and
$\mu$ simultaneously.)

One can use the results of Figs.~\ref{pt100contourplots} and
\ref{nlocontourplots} to estimate a theoretical error to be
ascribed to the calculation, as described in the introductory
paragraphs. We can estimate the error as the amount by which the cross
section changes when one makes a ``substantial'' change in $\mu$ and
$\lambda$. But we need an {\it a priori} decision about what range of
$\mu$ and $\lambda$ represents a substantial change. Evidently such a
decision must be subjective but we cannot avoid making a choice.

Consider $\mu$ first. This parameter is supposed to represent the
typical momentum that flows in loops of Feynman graphs for the
process. For instance, a good guess would be $\mu = E_T\, \sin(R)$
where $R = 0.7$ is the cone size. Thus $\mu = E_T/2$ is not an
unreasonable guess. Clearly such an estimate cannot be good to much
better than a factor of two, but our intuition is that the estimate
should not be off by much more than a factor of two either. Thus we
adopt a factor of two as a measure of a ``substantial'' scale change.

Now consider the parameter $\lambda$. The choice $\lambda = 0$
represents the \MSbar\ convention. The choice $\lambda = 1$ represents
the DIS convention, in which the contribution to deeply inelastic
scattering from gluon initial partons is canceled by choice of
convention. Since this represents a qualitative alteration in how
deeply inelastic scattering appears to happen, it seems to us that
$\Delta \lambda = 1$ represents a ``substantial'' change.

We thus propose that the variation in the cross section when $\mu$ and
$\lambda$ vary in the range $\{\Delta\log_2(\mu) = \pm 1,
\Delta \lambda = \pm 1\}$
represents a minimum theoretical error in the sense that it would be a
surprise (to us anyway) if the difference between the NLO result and a
NNLO result were less than that. A more conservative error estimate
would come from doubling the range: $\{\Delta\log_2(\mu) = \pm 2,
\Delta \lambda = \pm 2\}$.

With this understanding, we can read error estimates off of
Fig.~\ref{nlocontourplots}. The minimum error estimate varies from 3\%
at $E_T = 50 \GeV$ to 7\% at $E_T = 400 \GeV$. The conservative error
estimate varies from 9\% at $E_T = 50 \GeV$ to 32\% at $E_T = 400
\GeV$.

\begin{figure}[h]
\centerline{\epsfbox{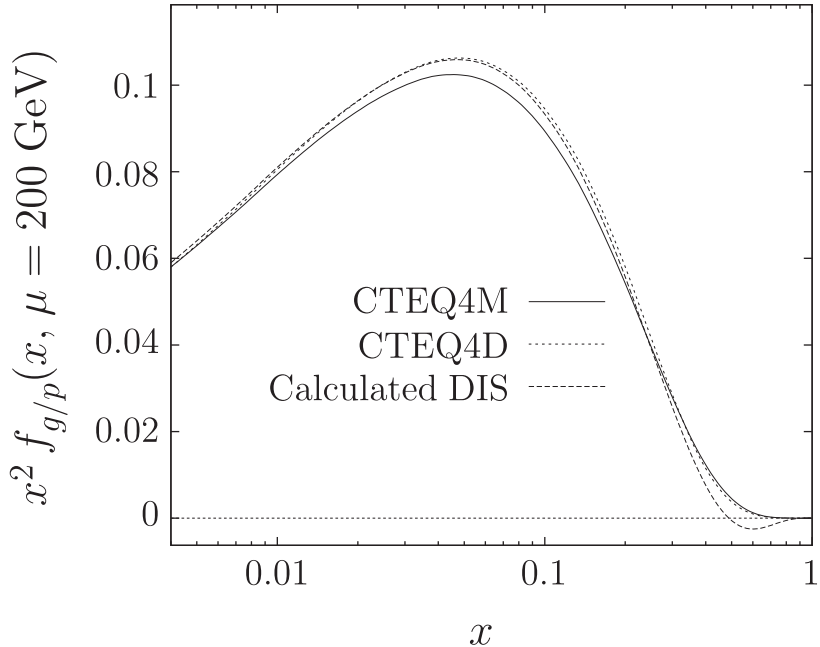}}
\vspace{0.2in}
\caption{CTEQ4M, CTEQ4D and calculated gluon distribution at 
$\mu = \mathrm{200 \ GeV}$} 
\label{gluonplot}
\end{figure}

Klasen and Kramer have carried out an interesting investigation
\cite{KK} that is in some respects similar to that reported here. They
calculated the jet cross section at large $E_T$ using the CTEQ3D
parton distributions, which are defined with the DIS convention. They
compared this cross section to the cross section calculated using the
CTEQ3M parton distributions, which are defined with the \MSbar\
convention. In each case, they used the $\hat\sigma$ appropriate to
the parton convention. They found a large difference. We agree with
this result. We have calculated $d\sigma/dE_T$ at $E_T = 400 \GeV$
with $\mu = E_T/2$ using CTEQ4D parton distributions and the DIS
version of $\hat\sigma$. We find that this DIS cross section is 33\%
greater than the corresponding \MSbar\ cross section calculated using
CTEQ4M parton distributions and the \MSbar\ version of $\hat\sigma$.
How can this be consistent with Fig.~\ref{nlocontourplots}, which
indicates that the DIS cross section ($\lambda = 1$) differs from the
\MSbar\ cross section ($\lambda = 0$) by less than 1\% for $\mu =
E_T/2$? The answer is that the DIS parton distributions obtained from
the CTEQ4M distributions by using the transformation (\ref{transform})
are not the same as the CTEQ4D parton distributions.

We can comment in more detail.  The CTEQ4D distributions are not
obtained from the CTEQ4M distributions by using Eq.~(\ref{transform})
but are instead obtained by fitting the world's data using a DIS
version of NLO theoretical formulas.  In Fig.~\ref{gluonplot} we
display the gluon distribution at $\mu = 200 \GeV$ as given by the
CTEQ4M set, by the CTEQ4D set, and by the calculated transformation
(\ref{transform}) from the CTEQ4M set.  We see that the DIS gluon
distribution calculated using Eq.~(\ref{transform}) is negative for $x
> 0.5$ for $\mu = 200 \GeV$. The CTEQ4D gluon distribution is positive
for all $x$. (This is a constraint imposed in the fitting procedure.)
Thus the CTEQ4D gluon distribution is substantially larger at large
$x$ than it might have been. This does not create a bad fit since
there is essentially no data that constrains the gluon distribution at
large $x$ \cite{CTEQglue}. The larger gluon distribution leads to a
larger jet cross section. Thus the 33\% change in the calculated cross
section can be attributed to the uncertainties in fitting parton
distributions, arising ultimately from the fact that the gluon
distribution at large $x$ is not constrained by data.

In summary, we have investigated a two dimensional space of
factorization schemes. The dependence of the jet cross section on the
two parameters $\mu$ and $\lambda$ is moderate in the range
$\mathrm{50 \ GeV} \le E_T \le \mathrm{400 \GeV}$. The sensitivity to
$\mu$ and $\lambda$ increases as $E_T$ increases. Within the space
investigated, the choice $\mu=E_T/2,\ \lambda=0$ (the \MSbar\ scheme
with a standard choice of scale) seems as good as other nearby
choices. It will be interesting to see whether these conclusions
change when we investigate a bigger space of factorization schemes.

\end{document}